\documentclass{evn2004}
\usepackage{txfonts}
\usepackage{graphicx}
\begin{document}
\title{Towards the Event Horizon - The Vicinity of AGN at Micro-Arcsecond Resolution}

\author{T.P. Krichbaum\inst{1}
\and D.A. Graham\inst{1}
\and W. Alef\inst{1}
\and A. Kraus\inst{1}
\and B.W. Sohn\inst{1}
\and U. Bach\inst{1}
\and A. Polatidis\inst{1}
\and A. Witzel\inst{1}
\and J.A. Zensus\inst{1}
\and M. Bremer\inst{2}
\and A. Greve\inst{2}
\and M. Grewing\inst{2}
\and S. Doeleman\inst{3}
\and R.B. Phillips\inst{3}
\and A.E.E. Rogers\inst{3}
\and H. Fagg\inst{4}
\and P.~Strittmatter\inst{4}
\and L. Ziurys\inst{4}
\and J. Conway\inst{5}
\and R. S. Booth\inst{5}
\and S. Urpo\inst{6}
}

\institute{Max-Planck-Institut f\"ur Radioastronomie, Bonn, Germany
\and Institut de Radioastronomie Millim\'etrique, Grenoble, France
\and MIT-Haystack Observatory, Westford, MA, USA
\and Steward Observatory, University of Arizona, Tucson, AZ, USA
\and Onsala Space Observatory, Onsala, Sweden
\and Mets\"ahovi Research Station, Helsinki University, Finland
}

   \abstract{
We summarize the present status of VLBI experiments at 3\,mm (86\,GHz), 2\,mm (129-150\,GHz)
and 1.3\,mm (215-230\,GHz). We present and discuss a new 3\,mm VLBI map of M\,87, which has a spatial resolution
of only $\sim 20$ Schwarzschild radii. We discuss recent results for Sgr\,A* and argue in favor of new
observations within an extended European mm-VLBI network, in order to search for variability. 
We discuss the possibilities to image the `event horizon' of a super-massive black hole at 
wavelengths $< 2$\,mm, and conclude that the addition of large and sensitive millimetre telescopes such 
as CARMA, the SMA, the LMT and ALMA will be crucial for this.
   }

   \maketitle
%

\section{Introduction}
Very Long Baseline Interferometry at millimetre wavelengths (mm-VLBI) allows the detailed imaging of
compact galactic and extragalactic radio sources with angular resolutions unreached by any other
astronomical observing technique. It offers a unique possibility to image
the direct vicinity of the super-massive black holes (SMBH) thought to be located
in the centres of powerful radio galaxies and other Active Galactic Nuclei (AGN), including
the SMBH in our Galaxy (Sgr\,A*). The highest possible
angular and spatial resolution is also required to answer the still unsolved and fundamental question of how
the powerful radio jets of AGN are launched and how they are accelerated and collimated.

In radio interferometry the angular resolution can be improved either by increasing
the separation between the radio telescopes, or by observing at shorter wavelengths. The first
possibility leads to VLBI with orbiting radio antennas in space (e.g. VSOP, ARISE). The second
alternative leads to mm-VLBI, in which ground-based radio telescopes observe at frequencies
above $\sim 80$\,GHz. The resolution of a global mm-VLBI array at 230 GHz would be about $25-30$
micro-arcseconds (1 $\mu$as = $10^{-6}$ arcsec), similar to the resolution of future space-VLBI at 
43 GHz. 
Since the innermost region of an AGN is invisible at centimetre wavelengths
(due to intrinsic self-absorption), mm-VLBI offers the additional advantage to penetrate
this opacity barrier, opening the direct view onto the ``central engine''.

Here we report on recent developments in mm-VLBI, with particular emphasis on
VLBI experiments performed at the highest accessible VLBI frequencies of 86, 150 and 230 GHz.
We demonstrate that global VLBI at 230 GHz now is technically feasible and yields detections of AGN with an
angular resolution of $\sim 30 \mu$as. When combined with future 
antennas like CARMA (USA), the SMA (Hawaii), the LMT (Mexico), and ALMA (Chile), the sensitivity
could be increased to a level so that detailed studies of galactic 
and extragalactic (super-massive) black holes with a spatial resolution of only a few Schwarzschild radii 
($R_{\rm S}$) will become possible.

\vspace{-0.5cm}
\section{Imaging the Base of the Jet in M\,87 with 20 ${\bf R_{\rm S}}$}
The Global 3\,mm-VLBI Array (GMVA) has been operational since 2003 (see http://www.mpifr-bonn.mpg.de/\-div/\-vlbi/\-globalmm).
At 86\,GHz, it combines the European antennas with the VLBA. In Europe, the following observatories
participate in the GMVA: Effelsberg, Onsala, Mets\"ahovi, Pico Veleta and Plateau de Bure. When 
compared with the stand-alone VLBA, the GMVA is a factor of $3-4$ more sensitive. This is mainly due to
the participation of the two IRAM telescopes (the 30\,m telescope on Pico Veleta in Spain, and the 6x15\,m
inter\-fero\-meter on Plateau de Bure in France). The GMVA is open to the scientific community and
presently observes twice per year, in spring and autumn. In each session and for logistical reasons, 
the observations are scheduled in time blocks of $3-6$ days duration, depending on proposal pressure. 
For the near future (2005 onwards) it is planned to change the default recording mode from presently 256 Mbit/s 
to 512 Mbit/s, giving better sensitivity.  Since at the VLBA the tape consumption is 
limited (to $\leq$ 2 tapes in 24 hrs), the duty cycle for the recording (= time used for recording / total time) 
is at present only 0.21 (for 512 Mbit/s), and 0.43 (for 256 Mbit/s). The duty cycle and uv-coverage could
be increased, if in the future the VLBA were able to change the tapes more frequently or if the VLBA would record
on hard-disks. The latter has the additional advantage that recording rates of $> 512$ Mbit/s would become 
possible also globally.

As an example for an image obtained with the Global 3\,mm-VLBI Array, we show in Figure 1 a new 86 GHz 
VLBI image of the inner jet of M\,87 (= 3C\,274). The following stations contributed to this map:
Effelsberg (B), Onsala (S), Pico Veleta (V), Plateau de Bure (phased array) (P), and 8 VLBA stations (all except BR and SC).
The data were recorded at 256 Mbit/s using the MK5 disk recording in Europe and tape recording
at the VLBA. The source was detected on the B-V-P baselines with SNR $\leq 200$ and on the baselines to and within the VLBA
with SNR $\leq 50$. After initial fringe fitting at the Bonn correlator and narrowing of the search windows using 
3C\,273 as fringe tracer (SNR $< 330$), the data were imported into AIPS with the new task 
MK4IN (Alef \& Graham 2002). The final fringe fitting and amplitude calibration was done using the 
standard procedures in AIPS. The final imaging was done using the Difmap package.

At a distance of 18.7 Mpc of M\,87, the angular resolution of 300 x 60 $\mu$as
corresponds to a spatial scale of 30 x 6 light days, or 100 x 20 Schwarzschild radii (assuming a
3 x $10^9$ $\rm{M}_\odot$ BH). Thus, the central engine and the inner jet can be studied  
with a similar spatial resolution as the less massive, but closer SMBH in Sgr\,A*.
In fact, owing to its higher declination, M\,87 is easier to observe with VLBI and might be
even a better candidate than Sgr\,A* for the imaging of the event horizon around a SMBH.
One of the main differences between these two objects of course is, that the jet of M\,87 is related to a radio-loud 
galaxy, whereas Sgr\,A* has a much lower radio-luminosity and shows no jet. The study
of both sources therefore should help to obtain a better understanding how jets are formed in general, and 
how they are accelerated and collimated.
The fact that in M\,87 the jet can be traced down to scales of only a few ten Schwarzschild radii without
a large reduction of its brightness temperature is very noteworthy. This may
give new constraints to the theories of jet formation. 
The comparison of the width of the jet at its origin with the expected size of the light cylinder
can help to discriminate between jet models, e.g. whether magnetic sling-shot models (e.g. Blandford \& Payne 1982) or
models with direct coupling to the BH spin (e.g. Blandford \& Znajek 1977) are more appropriate.
\begin{figure}
\includegraphics[bb=50 185 470 625,clip=,angle=-90,width=.5\textwidth]{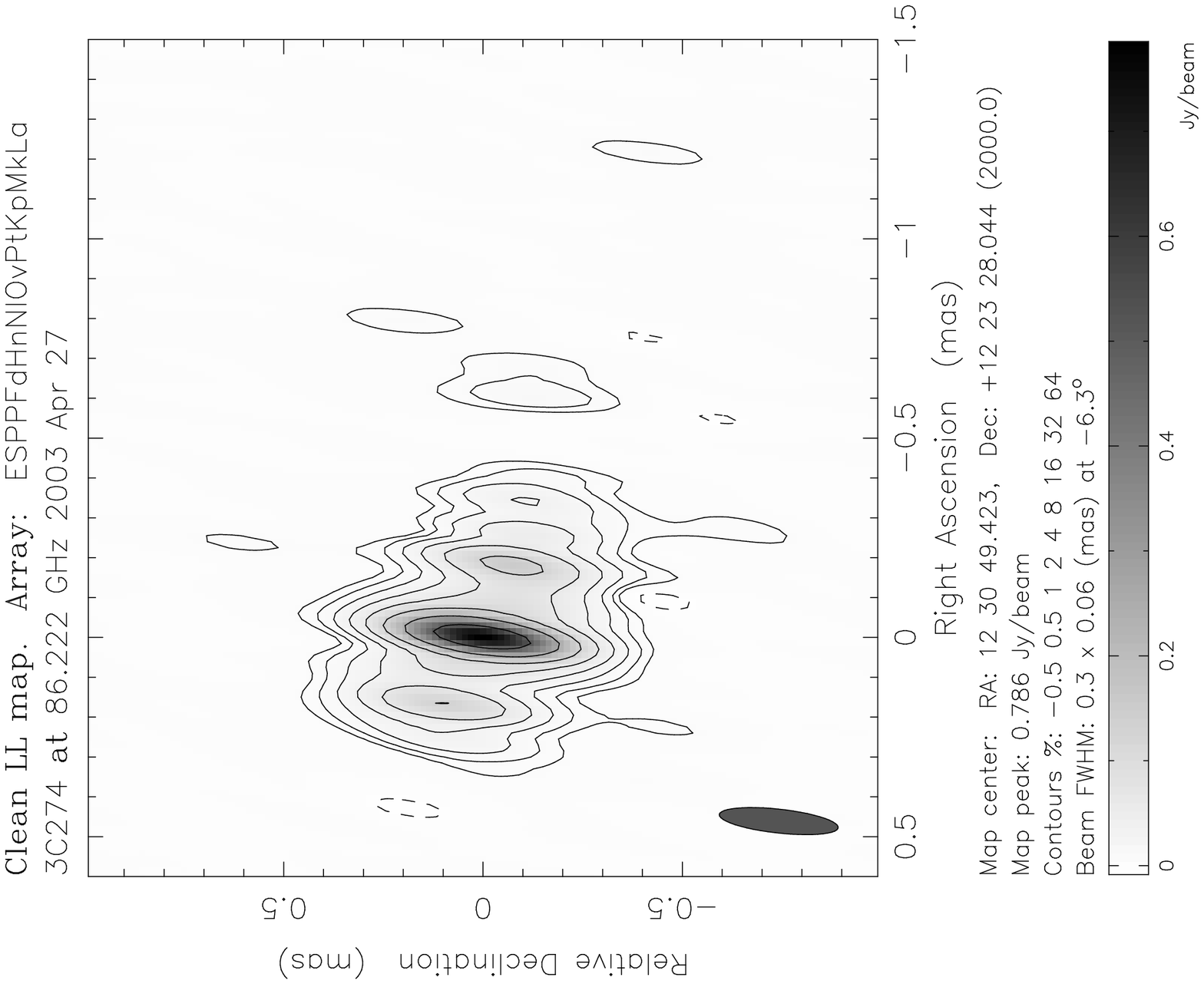}
\caption{ 
VLBI image of M\,87 (3C\,274, Virgo A) obtained in April 2003 at 86 GHz with the Global Millimetre VLBI Array (GMVA).
Contour levels are -0.5, 0.5, 1, 2, 4, 8, 16, 32, and 64 \% of the peak flux of 0.79 Jy/beam.
The beam size is 0.30 x 0.06 mas, pa=-6.3$^\circ$. The identification of the easternmost
jet component as VLBI core or as a counter-jet is still uncertain.}
\end{figure}

\vspace{-0.3cm}
\subsection{VLBI Observations of Sgr\,A* with European Antennas}
\begin{table}
\begin{tabular}{llclcc}
Station     &Country  &Diameter & T$_{\rm sys}^{\rm zenith}$  &Efficiency & SEFD \\ 
            &         &[m]      & [K]   &    [\%]       & [Jy] \\ \hline
Effelsberg  &Germany  & 100     &  130  &    8      &  950 \\
Pico Veleta &Spain    &  30     &  120  &   55      &  860  \\
P. de Bure  &France   & 6x15    &  120  &   65      &  570  \\
Onsala      &Sweden   & 20      &  200  &   45      & 3900 \\
Mets\"ahovi &Finland  & 14      &  200  &   30      & 12000 \\ \hline
Yebes       &Spain    & 40      &  150  & $\sim$40      &  $\sim$830  \\
Noto        &Italy    & 32      &  150  & $\sim$30      &  $\sim$2300  \\
SRT         &Italy    & 64      &  150  & $\sim$40      &  $\sim$430  \\
\end{tabular}

\caption{Present and future telescopes, which in principle could enlarge the
European VLBI array at 3\,mm (80-90 GHz). The station names and locations are shown in columns 1 \& 2. Column 3
gives the antenna diameter, column 4 a typical system temperature (at zenith) of the receiver, column 5 the aperture
efficiency and column 6 the system equivalent flux density (SEFD) defined as the ratio of T$_{\rm sys}$ and
antenna gain (in [K/Jy]). The top 5 stations in the table participate regularly in global 3mm-VLBI observations 
together with the VLBA.  The stations listed below the intersecting line are future candidates. 
}
\vspace{-0.5cm}
\end{table}
The compact radio source Sgr\,A*, which is located at the Centre of our Galaxy, most likely harbors
the nearest super-massive black hole. VLBI-images of Sgr\,A* at cm-wavelengths are heavily
affected by interstellar scatter broadening (e.g. Marcaide et al. 1999 and references therein), which blurs
the underlying source structure. Since the image broadening decreases
quadratically with increasing frequency, VLBI observations at mm-wavelengths allows us to
penetrate the scattering screen and image the source behind. Early VLBI observations
at 7\,mm (Krichbaum et al. 1993, Lo et al. 1998), 3\,mm (Krichbaum et al. 1998, Doeleman et al. 2001) and 1\,mm 
(Krichbaum et al. 1998)
already indicated that the source appears slightly larger than the expected scattering size. New VLBA
observations at 7\,mm confirm this effect and now also suggest a possible variation of
the VLBI structure with time (Bower et al. 2004). Structural variability is not unexpected in view of the flux density
variations seen in the radio and near infrared bands (e.g. Zhao et al. 2001, Genzel et al. 2004),
and may provide an important building block for our understanding of the true nature of this enigmatic source.

Motivated by this, we simulate 3\,mm VLBI observations of Sgr\,A* with the existing European antennas and adding 
new telescopes which may be able in the near future to participate in mm-VLBI. Here we argue that it is 
a worthwhile effort to equip the following stations with 3\,mm receivers: the new 40\,m antenna built
by the Yebes group, the 32\,m antenna with its new adaptive reflector in Noto (Sicily), and the 64\,m Sardinia
radio telescope. In Table 1 we summarize the antenna characteristics, in Figure 2 we show uv-coverages for
Sgr\,A*, subsequently adding the new stations (note: Sgr\,A* is invisible from  Onsala and 
Mets\"ahovi). 

With the VLBA, 86\,GHz VLBI observations of Sgr\,A* theoretically should yield an image with a maximum 
resolution of 0.07\,mas (minor axis of beam). When the partially resolved source flux falls below 
the baseline detection threshold, the resolution degrades. With a size of 0.18\,mas for Sgr\,A* at 86\,GHz
and 0.4-0.5\,Jy baseline sensitivity (assuming 512 Mbit/s, SEFD = 4300 Jy), the resulting maximum angular 
resolution of the VLBA is 650\,M$\lambda$ or 0.16\,mas. 

An European array with the telescopes listed in Table 1 could image the source with quite similar
resolution (500 \,M$\lambda$, 0.21\,mas) than the VLBA, but also with higher sensitivity. While the source would be
just marginally detected on the 500-600\,M$\lambda$ VLBA baseline with a SNR of $5-7$, 
the sensitive European baselines (e.g. Effelsberg to IRAM) would see it with a SNR of $20-45$ 
(baseline detection thresholds: VLBA - VLBA : 480\,mJy, Pico - Effelsberg: 100\,mJy; Pico - PdBure: 76\,mJy).
When combined with other European mm-VLBI stations located in southern Europe (see in Fig. 2), very good
3\,mm-VLBI images of Sgr\,A* could be obtained. The high SNR of the measured visibilities and the good
uv-coverage would facilitate a more accurate determination of the source size and structure, with smaller 
uncertainties than in previous VLBI observations. Small error bars on the source size, however, are 
absolutely necessary for the clear detection of structural variability. 

We conclude: the large and sensitive new radio telescopes being built in Spain and Italy, will
significantly extend the European VLBI baselines to the south. This will result in 
better VLBI images of many radio sources, particularly for those with relatively low 
declinations (e.g. M\,87).  At short cm- and mm-wavelengths, where the
uv-coverage of the existing arrays (EVN: at 1.3 \& 7\,mm, GMVA: at 3\,mm) is still not very dense, the
participation of these new telescopes would lead to much better VLBI images. 
If equipped with 3\,mm receivers, these stations could also play an important role in 
the imaging of nearby SMBHs, as e.g. for Sgr\,A*.
\begin{figure}
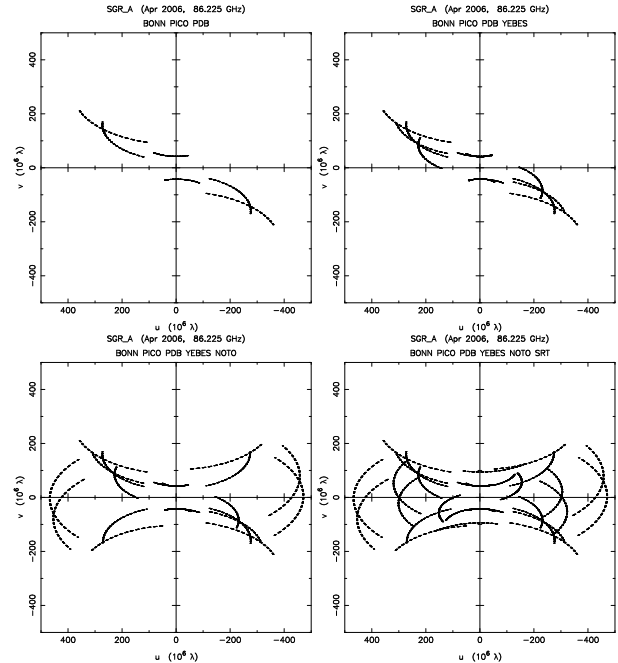

\includegraphics[width=.22\textwidth]{krichbaum-fig2a.ps}
\includegraphics[width=.22\textwidth]{krichbaum-fig2b.ps}

\includegraphics[width=.22\textwidth]{krichbaum-fig2c.ps}
\includegraphics[width=.22\textwidth]{krichbaum-fig2d.ps}

\caption{Simulated uv-coverages for a VLBI observation of Sgr\,A* at 86\,GHz.
The simulations are done for the following telescopes (4 simulations arranged
from top left to bottom right): (a) Effelsberg, Pico Veleta, Plateau de Bure (present array),
(b) plus Yebes, (c) plus Noto, (d) plus SRT (Sardinia Radio Telescope).
}
\vspace{-0.5cm}
\end{figure}

\section{Towards Shorter Wavelengths - VLBI at 2\,mm and 1\,mm}
In order to demonstrate the technical feasibility of VLBI at wavelengths shorter than 3\,mm,
several VLBI pilot experiments were performed. At 2\,mm (147\,GHz) the following
telescopes were available: Pico Veleta (30\,m, Spain), Heinrich-Hertz Telescope (10\,m, Mt. Graham, Arizona), Kitt Peak
Telescope (12\,m, Kitt Peak, Arizona), Mets\"ahovi (14\,m, Finland) and SEST (15\,m, Chile). In two experiments performed
in 2001 and 2002, about one dozen mm-bright quasars were detected on the short continental baselines
in Europe (Pico-Metsa) and in the USA (HHT-KP) (Greve et al. 2002, Krichbaum et al. 2002).
A big success was the detection of 3 quasars also on the 4.2\,G$\lambda$ long transatlantic baseline between Pico Veleta
and the Heinrich-Hertz Telescope: NRAO150 (SNR=7), 1633+382 (SNR=23) and 3C279 (SNR=75).
In addition to continuum sources at 147\,GHz, also several SiO masers were observed (at 129\,GHz) and detected
on short baselines (Doeleman et al. 2002).
This success motivated another VLBI experiment one year later (April 2003);
this time at the shorter wavelength of 1.3\,mm (230\,GHz). In this experiment
the following stations participated: Pico Veleta, the 6x15\,m IRAM interferometer on 
Plateau de Bure (as phased array), the Heinrich-Hertz Telescope and the 12\,m telescope on Kitt Peak.
Instead of recording on tapes, the new MK5 disk recording was chosen. The data were
recorded at a rate of 512 Mbit/s. In this observation, 
the following sources were detected on the 880\,M$\lambda$ long baseline between Pico Veleta and
Plateau de Bure: NRAO\,150 (SNR=10.7), 3C\,120 (SNR=8.2), 0420-014 (SNR=24.9), 0736+017 (SNR=7.1), 0716+714
(SNR=6.8), OJ287 (SNR=10.4), 3C\,273 (SNR=8.2), 3C\,279 (SNR=9.6), and BL\,Lac (SNR=9.0).
Sensitivity limitations and some technical problems restricted the number of detected sources
on the 6.4\,G$\lambda$ long transatlantic baseline between Pico Veleta and HHT to
the quasar 3C\,454.3 (SNR=7.3). The BL\,Lac object 0716+714 was marginally detected (SNR=6.4). 
No transatlantic fringes were seen to Plateau de Bure.
After the experiment and during correlation, it became obvious that the phase stability
of Plateau de Bure was not perfect and that some additional phase noise in the data degraded
the SNR of the detections on the baselines to this station by about a factor of 3-4. The problem is 
under investigation and will be fixed soon.

Although the number of sources detected on the Pico Veleta - HHT baseline still is small,
the results demonstrate the technical feasibility of global 1\,mm VLBI. The detections also
mark a new record in angular resolution in astronomy (size $< 32 \mu$as) and indicate
the presence of ultra-compact emission regions in AGN, even at the highest frequencies.
For the  quasar 3C\,454.3 (z=0.859, see also Pagels et al., this conference), the detection 
was made at a rest frame frequency of 428 GHz.
At 2 and 1.3\,mm-wavelengths, the brightness temperatures of the detected AGN appears not to be 
significantly lower than at cm-wavelengths. There are, however, indications that the source 
compactness might be variable (for different sources and for a given source also with time). This is not
unexpected considering the known and often dramatic flux density and spectral variability in quasars,
which is much more pronounced at mm- than at cm-wavelengths.

\section{Future Outlook}
Micro-arsecond resolution imaging of compact radio sources with mm-VLBI is now possible, but still needs
further improvement. To obtain
an image fidelity comparable to present day cm-VLBI images, one needs a better
uv-coverage and a lower single baseline detection threshold, i.e. a higher array sensitivity. 
The capabilities of global 3\,mm VLBI can be further improved by the addition of large telescopes
in Europe (Yebes, SRT), in the USA (GBT, CARMA) and in Central and South America (LMT, ALMA), even 
if not all of these telescopes are optimized for 3\,mm-VLBI. When compared to the stand-alone VLBA, a sensitivity
improvement by at least a factor of $5-10$ appears possible.

At the shorter wavelengths (2\,mm, 1\,mm) several bright sources are already detected on long transatlantic
baselines. This demonstrates the feasibility of VLBI at these short wavelengths. However the number of
available antennas still is very small and the uv-coverage therefore correspondingly sparse. Thus, the 
future success of VLBI at and below 2\,mm will depend critically on the availability of 
a larger number of mm-antennas, which can observe at $\lambda \leq  2$\,mm. Major
steps towards better sensitivity and uv-coverage would be the addition of
the LMT (in Mexico), and the addition of the large millimetre interferometers CARMA (in California),
the SMA (in Hawaii), and ALMA (in Chile) to the mm-VLBI array. In combination with these very sensitives telescopes, the
smaller millimetre telescopes (APEX, KP-12m, JCMT, HHT) would efficiently contribute to the global uv-coverage. 
One should also consider the ALMA prototype antennas, which are presently located in Socorro (New Mexico).
Their relocation to suitable places could fill existing gaps in the uv-plane. Only the combination of the 
large with the smaller mm- and sub-mm antennas will lead to a global mm-VLBI array, which finally has a
high enough sensitivity to allow the imaging of those regions, where the coupling between 
accretion disk, black hole and jet occurs.

In nearby objects and with a spatial resolution of only a few Schwarzschild radii, it should 
be possible to reach the "event horizon" or at least the inner part of the accretion disk around 
the central SMBH, if it radiates and is visible at mm-/sub-mm wavelengths.
Another important aspect and future goal is the VLBI polarimetry in the millimetre and sub-millimetre
domaine. At these short wavelengths, the jet base should become optically thin and the polarization should
be high. Sub-mm VLBI polarimetry therefore should allow observing the expected time-variable magnetic 
field configuration in the BH-jet system. This will facilitate detailed tests of relativistic magneto-hydrodynamical 
jet- and dynamo-models, which are presently proposed as a likely mechanism for jet creation. 

The ongoing development of the VLBI recording systems towards higher sampling rates and larger bandwidths 
(several Gbit/s) already points in the right direction and towards higher baseline sensitivities
(Graham et al., 2002, Whitney et al. 2003). At mm- and sub-mm
wavelengths, it will be also very important to correct instantaneously for the phase fluctuations introduced by the
Earth's atmosphere on short timescales (seconds to minutes). Simultaneous dual-frequency observations and/or 
water vapor radiometry will help to extend the phase coherence and integration times and by this contribute to the 
necessary sensitivity enhancement.
Thus one can hope that within less than a decade from now, the detailed imaging of the direct vicinity
of SMBHs and their `event horizon' will really become possible.

\begin{acknowledgements}
2\,mm and 1\,mm VLBI is a joint effort of the following observatories: MPIfR (Bonn), IRAM (France and Spain),
Mets\"ahovi Radio Observatory (Finland), MIT-Haystack Observatory (USA), Arizona Radio Observatory (USA),
and Onsala Radio Observatory (Sweden and Chile).
The results presented here would have been not possible without the help of many people at each of the
participating observatories. Thanks to them all!
\end{acknowledgements}

\end{document}